\documentclass[useAMS,usenatbib]{mn2e}
\usepackage{times,graphicx}
\usepackage{epsfig,wrapfig}
\usepackage{mathrsfs}
\usepackage{color}
\usepackage{epsfig,wrapfig}
\usepackage{type1cm}

\newcommand{\boldsymbol}[1]{\mbox{\boldmath $#1$}}

\newcommand{\degree}{\hbox{$^\circ$$$}}

\renewcommand{\theequation}{\textcolor{blue}{\arabic{equation}}}
\renewcommand{\thefigure}{\textcolor{red}{\arabic{figure}}}
\renewcommand{\thetable}{\textcolor{red}{\Roman{table}}}

\title[\textbf{On the Perihelion Precession \textit{of} Planetary Orbits}]{ Azimuthally Symmetric Theory \textit{of} Gravitation (I) \\
{\Large On the Perihelion Precession \textit{of} Planetary Orbits}}
\author[G. G. Nyambuya ]{G. G. Nyambuya\thanks{Email: gadzirai@gmail.com } \\
}

\begin{document}

\date{Received: 5 Sept. 2009 / Accepted with Moderate Revision: 9 Dec. 2009 }

\volume{0000}
\pagerange{\pageref{firstpage}--\pageref{lastpage}} \pubyear{2009}

\label{firstpage}

\maketitle

\begin{abstract}
{From a purely none-general relativistic standpoint, we solve the empty space Poisson equation ($\nabla^{2}\Phi=0$) for an azimuthally symmetric setting, \textit{i.e.}, for a spinning gravitational system like the Sun. We seek the general solution of the form $\Phi=\Phi(r,\theta)$. This general solution is constrained such that in the zeroth order approximation it reduces to Newton's well known inverse square law of gravitation. For this general solution, it is seen that it has implications on the orbits of test bodies  in the gravitational field of this spinning body. We show that to second order approximation, this azimuthally symmetric gravitational field is capable of explaining at least two things (1) the observed perihelion shift of solar planets (2) that the \textit{mean Earth-Sun distance} must be increasing -- this resonates with the observations of two independent groups  of astronomers (\citealt{krasinsky04}; \citealt{standish05})  who have measured that the \textit{mean Earth-Sun distance} must be increasing at a rate of about $7.0\pm0.2\, m/cy$ (\citealt{standish05}) to $15.0\pm0.3\, m/cy$ (\citealt{krasinsky04}). In-principle, we are able to explain this result as a consequence of loss of orbital angular momentum -- this loss of orbital angular momentum is a direct prediction of the theory. Further, we show that the theory is able to explain at a satisfactory level the observed secular increase Earth Year ($1.70\pm0.05\,ms/yr$; \citealt{miura09}). Furthermore, we show that the theory makes a significant and testable prediction to the effect that the period of the solar spin must be decreasing at a rate of at least $8.00\pm2.00\,s/cy$. }
{}
{}
{}
\end{abstract}

\begin{keywords}
astronomical unit, azimuthal symmetry, orbit, perihelion shift, solar spin
\end{keywords}

\section{ Introduction}

From as way back as the $1850s$, it has been known that the orbit of the planet Mercury exhibits a peculiar motion of its perihelion, specifically, the perihelion of Mercury advances by ${43.1}\pm{0.5}$ $\rm{arcsec/century}$. When Newton's theory of gravitation is applied to try and explain this (by making use of the oblateness of the planets because when the Sun's gravitational force acts on the oblate-planets, the oblateness causes torque [on the planets] and this torque is thought to give rise to the anomalous motion of the planets); it was found first by Leverrier in ${1859}$ see \textit{e.g.} \cite{kenyon90} that it predicted a precession of ${532}$ $\rm{arcsec/century}$ which is larger than the observed \citep{kenyon90}. With the failure of Newton's theory to explain this, it was proposed that a small undetected planet was the cause. Careful scrutiny of the terrestrial heavens by telescopes and spaces probes reveals no such object -- the meaning of which is that the cause may very well be a hitherto unknown gravitational phenomena -- Einstein was to  demonstrate that this was the case, that there existed a hitherto unknown gravitational phenomena that is the cause of this peculiar motion. 

With the herald of Einstein's General Theory of Relativity (GTR) in ${1915}$, Einstein immediately applied his GTR to this problem; much to his elation which caused him heart palpitations -- he obtained the unprecedented value of ${43.0}\,\rm{arcsec/century}$  and this was (and is still) hailed as one of the greatest triumphs for the GTR and this lead to its quick acceptance.  Venus, the Earth, and other planets show such peculiar motion of their perihelion. Observations reveal a shift of ${8.40}\pm{4.80}$ and ${5.00}\pm {1.00}\,\rm{arcsec/century}$ respectively (see \textit{e.g.} {Kenyon ${1994}$}). Einstein's theory is able to explain the perihelion shift of the other planets well, so much that it is now a well accepted paradigm that the perihelion shift of planetary orbits is a general relativistic phenomena.

Einstein's GTR explains the perihelion shift of planetary orbits as a result of the curvature of spacetime around the Sun. It does not take into account the spin of the Sun and at the same time it assumes all the planets lay on the same plane. The assumption that the planets lay on the same plane is in the GTR solution only taken as a first order approximation -- in reality, planets do not lay on the same plane. In this reading we set forth what we believe is a new paradigm; we have coined this paradigm the  Azimuthally Symmetric Theory \textit{of} Gravitation (ASTG) and this is derived from Poisson's well accepted equation for empty space -- namely $\nabla^{2}\Phi=0$. Poisson's Law is a differential form of Newton's Law \textit{of} Gravitation. We explain the perihelion shift of the orbits of planets as a consequence of the spin of the Sun -- \textit{i.e.} solar spin. It is well known that the Sun does exhibit some spin angular momentum -- specifically, it [the Sun] undergoes differential rotation. On the average, it spins on its spin axis about once in every $\sim25.38$ \textit{days} (see \textit{e.g.} \citealt{miura09}). Its spin axis makes an angle of about $83\degree$ with the ecliptic plane. It is important that we state clearly here that by no means have we discovered a new theory or a set of new equations; we have merely applied Poisson's well known azimuthally symmetric solution to gravity for a spinning gravitating body.

\begin{figure}
\centering
\label{scoord}
\includegraphics[scale=0.8]{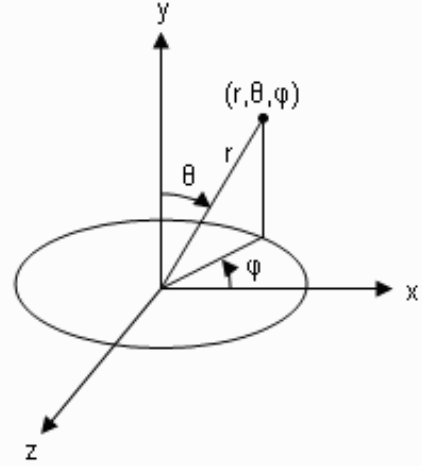}
\caption[\textbf{Generic Spherical Coordinate System}]{This figure shows a generic spherical coordinate system, with the radial coordinate denoted by $r$, the zenith (the angle from the North Pole; the colatitude) denoted by $\theta$, and the azimuth (the angle in the equatorial plane; the longitude) by $\varphi$.}
\end{figure}

Further, with regard to Einstein's GTR -- \textit{vis}; in its solution  to the problem of the perihelion shift of planetary orbits, it [the GTR] assumes the traditional Newtonian gravitational potential, namely: $\Phi(r)=-G\mathcal{M}/r$, where $G=6.667\times10^{-11}kg^{-1}ms^{-2}$ is Newton's universal constant of gravitation, $\mathcal{M}$ is the mass of the central gravitating body and $r$ is the radial distance from this gravitating body.  Einstein's GTR which is embodied in Einstein's law of gravitation, namely:

\begin{equation}
R_{\mu\nu}-\frac{1}{2}Rg_{\mu\nu}=\kappa T_{\mu\nu}+\Lambda g_{\mu\nu},
\end{equation}

is designed such that in the low energy limit and low spacetime curvature such as in the Solar System, this equation reduces directly to Poison's equation. In Einstein's law above, $R_{\mu\nu}$ is the Ricci tensor, $R$ the Ricci scalar, $g_{\mu\nu}$ the metric of spacetime, $\Lambda$ is Einstein's controversial cosmological constant which at best can be taken to be zero unless one is making computations of a cosmological nature where darkenergy is involved, and $\kappa=8\pi G/c^{4}$  where $c=2.99792458\times10^{8}ms^{-1}$ is the speed of light in vacuum; and Poisson's equation is given by:

\begin{equation}
\nabla^{2}\Phi=4\pi G\rho\label{poison},
\end{equation}

where $\rho$ is the density of matter and the operator $\vec{\nabla}^{2}$ written for spherical coordinate system (see figure \ref{scoord} for the coordinate setup) is given by:

\begin{equation}
\vec{\nabla}^{2}= \frac{1}{r^{2}}\frac{\partial}{\partial r}\left(r^{2}\frac{\partial}{\partial r}\right)+\frac{1}{r^{2}\sin\theta}\frac{\partial}{\partial \theta}\left(\sin\theta\frac{\partial}{\partial\theta}\right)+\frac{1}{r^{2}\sin^{2}\theta}\frac{\partial^{2}}{\partial \varphi^{2}}.
\end{equation}

As already been said, our solution or paradigm, hails directly from Poisson's equation, which in itself is a first order approximate solution to Einstein's GTR, albeit with the important difference that we have taken into account solar spin. This fact that our paradigm explains reasonably well -- within the confines of its error margins; the precession of planetary orbits as a consequence of solar spin and at the sametime the GTR explains this same phenomena well as a consequence of the curvature of spacetime raises the question {\textbf{``Is the precession of the perihelion of solar orbits a result of (1) solar spin or (2) is it a result of the curvature of spacetime?''}} If anything, this is the question that this reading seems to raise. An answer to it, will only come once the meaning of the ASTG is fully understood. 

In the above we say the ASTG \textit{``explains reasonably well -- within the confines of its error margins''} -- what immediately comes to mind is that can a theory have error margins or is it not experiments that have error margins? As will be seen, certain undetermined constants ($\lambda_{\ell}$) in the theory emerge and at present, one has to infer these from observations and it is here that the error margins of the ASTG come into play. 

Further, we show, that \textit{in-principle}, the ASTG does explain (1) the increase in the mean Earth-Sun distance, (2) the increase in the mean Earth-Moon distance \textit{etc}, and these emerge as a consequence of the fact that from the ASTG, the orbital angular momentum is not a conserved quantity as is the case in Newtonian's gravitational theory and Einstein's  GTR. That the orbital angular momentum is not a conserved quantity may lead one to think that the ASTG violets the Law \textit{of} Conservation of angular momentum -- no, this is not the case. The lost angular momentum is transferred to the spin of the orbiting body and as well as the Sun.

\section{Theory}

For empty space: $\nabla^{2}\Phi=0$; and for a spherically symmetric setting we have $\Phi=\Phi(r)$ and this leads directly to Newtonian gravitation. For a scenario or setting that exhibits azimuthal symmetry such as a spinning gravitating body as the Sun we must have: $\Phi=\Phi(r,\theta)$, we thus shall solve the Poisson equation: $\nabla^{2}\Phi(r,\theta)=0$. The Poisson equation for this setting is readily soluble and its solution can readily be found in most of the good textbooks of electrodynamics and quantum mechanics for example -- it is instructive that we present this solution here. 

We shall solve Poisson's equation for empty space ($\nabla^{2}\Phi=0$) exactly; by means of separation of variables, \textit{i.e.} we shall set: $\Phi(r,\theta)=\Phi(r)\Phi(\theta)$. Inserting this into the Poisson equation we will have after some basic algebraic operations:

\begin{equation}
\frac{1}{\Phi(r)}\frac{\partial}{\partial r}\left(r^{2}\frac{\partial \Phi(r)}{\partial r}\right)+\frac{1}{\Phi(\theta)}\frac{1}{\sin\theta}\frac{\partial}{\partial \theta}\left(\sin\theta\frac{\partial\Phi(\theta)}{\partial\theta}\right)={0}.
\end{equation}

The radial and the angular portions of this equation must equal some constant since they are independent of each other. Following tradition, we must set:

\begin{equation}
\frac{1}{\Phi(r)}\frac{\partial}{\partial r}\left(r^{2}\frac{\partial \Phi(r)}{\partial r}\right)=\ell(\ell+1),
\end{equation}

and the solution to this is:

\begin{equation}
\Phi_{\ell}(r)= A_{\ell}r^{\ell}+\frac{B_{\ell}}{r^{\ell+1}}, 
\end{equation}

where $A_{\ell}$ and $B_{\ell}$ are constants and $\ell={0,1,2,3}, ...$ . If we set the boundary conditions; $\Phi_{\ell}(r=\infty)={0}$, then $A_{\ell}={0}$ for all $\ell$.  Now, just as Einstein demanded of his GTR to reduce to the well known Poisson equation in the low energy regime of minute curvature, we must demand that $\Phi(r)$,  in its zeroth order approximation -- where $\ell={0}$ and the terms $\ell\geq{1}$ are so small that they can be neglected; the theory must reduce to Newton's inverse square law; for this to be so,  we must have:
 
\begin{equation}
B_{\ell}=-\lambda_{\ell}c^{2}\left(\frac{G\mathcal{M}}{c^{2}}\right)^{\ell+1}, \label{bdef}
\end{equation}

where $\lambda_{\ell}$ is an infinite set of dimensionless parameters such that $\lambda_{0}=1$ and the rest of the parameters $\lambda_{\ell}$ for $\ell>1$ will take values different from unity and these constants will have -- for now, until such a time that we are able to deduce them directly from theory; to be determined from the experience of observations. In the discussion section, we shall hint at our current thinking on the nature of these constants. With this given, it means we will have:

\begin{equation}
\label{gen-pot}
\Phi_{\ell}(r)=  -\lambda_{\ell} c^{2}\left(\frac{G\mathcal{M}}{rc^{2}}\right)^{\ell+1}.\label{pdef}
\end{equation}

Now, moving onto the angular part, we will have:

\begin{equation}
\frac{\sin\theta}{\Phi(\theta)}\frac{\partial}{\partial \theta}\left(\sin\theta\frac{\partial\Phi(\theta)}{\partial\theta}\right)+\left[\ell(\ell+{1})\right]\sin^{2}\theta={0},\label{theta-comp}
\end{equation}

and a solution to this is a little complicated; it is given by the spherical harmonic function:

\begin{equation}
\Phi(\theta)=P_{\ell}(\cos\theta),
\end{equation}

of degree $\ell$ and $P_{l}(\cos\theta)$ is associated Legendre polynomial. As already said, the derivation of $\Phi(r,\theta)$ just presented can be found in most good standard textbooks of quantum mechanics and classical electrodynamics. Since equation (\ref{theta-comp}) is a second order differential equation, one would naturally expect there to exist two independent solutions for every $\ell$. It so happens that the other solutions give infinity at $\theta=({0},\pi)$, which is physically meaningless (see \textit{e.g.} Grifitts $2008$).  Now, putting all the things together, the most general solution is given:

\begin{equation}
\Phi(r,\theta)=-\sum^{\infty}_{\ell=0}\left[ \lambda_{\ell}c^{2}\left(\frac{G\mathcal{M}}{rc^{2}}\right)^{\ell+1}P_{l}(\cos\theta)\right],\label{newpot}
\end{equation}

which is  a linear combination of all the solutions for $\ell$. In the case of ordinary bodies such as the Sun,  the higher orders  terms [\textit{i.e.} $\ell>{1}$: of the term $(G\mathcal{M}/rc^{2})^{\ell+1}$],  will be small and in these cases, the gravitational field will tend to Newton's gravitational theory. Equation (\ref{newpot}) is the embodiment of the ASTG, and from this, we shall show that one is able to explain the precession of the perihelion of planetary orbits. 

In this equation [\textit{i.e.}, (\ref{newpot})], nowhere does the value of the Sun's spin ($\mathcal{T}_{\tiny \odot}\simeq25.38$ \textit{days}) enter into our equation. This may lead one to asking ``So where has this been taken into account?''.  To answer this, it is important to note that if the potential is a function of $r$ only \textit{i.e.}, $\Phi=\Phi(r)$, then, it technically is a function of $r$ and $\theta$ as well (with the $\theta$-dependence being trivial). What this means is that spherical symmetry implies an azimuthal symmetry around any arbitrarily chosen axis. If a specific axis is singled out, \textit{e.g.}, by the spin of a body about the spin axis, then, the spherical symmetry of the static body is broken, and only an azimuthal symmetry remains and this azimuthal symmetry is only about the plane cutting the body into hemispheres such that this plane is normal to the spin axis. For any other plane cutting the body into hemispheres, the two hemispheres are asymmetric. From this we see that the azimuthally symmetric solution is consequence of  the breaking of the spherical symmetry by the introduction of a spin axis,  hence thus one is automatically lead  to consider the solutions for which $\Phi=\Phi(r,\theta)$. In this way, the spin has been taken into account.

\begin{figure}
\centering
\label{pshift-fig}
\includegraphics[scale=0.30]{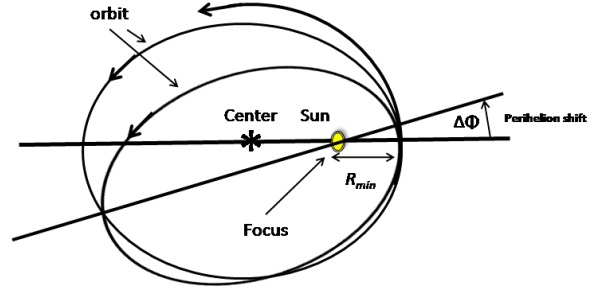}
\caption[\textbf{Perihelion Shift of Planets}]{The elliptical planetary orbits have the Sun at one focus. As the planets describe their orbits, their major axes slowly rotate about the Sun in the process shifting the line from the Sun to the perihelion through an angle $\Delta \varphi$ during each orbit. This shift is referred to as the precession of the perihelion.}
\end{figure}

\subsection{Equations of Motion}

We shall derive here the equations of motion for the azimuthally symmetric gravitational field, $\Phi(r,\theta)$. We know that the force per unit mass [or the acceleration \textit{i.e.}, $\vec{\textbf{g}}=-\nabla\Phi(r,\theta)$] is given by $\vec{\textbf{a}}=(\ddot{r}-r\dot{\varphi}^{2})\hat{\textbf{r}}+(r\ddot{\varphi}+2\dot{r}\dot{\varphi})\hat{\boldsymbol{\theta}}$ (see any good textbook on Classical Mechanics) where a single dot represents the time derivative $d/dt$ and likewise a double dot presents the second time derivative $d^{2}/dt^{2}$. Comparison of $\vec{\textbf{a}}=(\ddot{r}-r\dot{\varphi}^{2})\hat{\textbf{r}}+(r\ddot{\varphi}+2\dot{r}\dot{\varphi})\hat{\boldsymbol{\theta}}$ with ($\vec{\textbf{g}}$); \textit{i.e.}: $\vec{\textbf{a}}\equiv\vec{\textbf{g}}$, leads to the equations:

\begin{equation}
\frac{d^{2}r}{dt^{2}}-r\left(\frac{d\varphi}{dt}\right)^{2}=-\frac{d\Phi}{dr},\label{eqnmotion1}
\end{equation}

for the $\hat{\bf{r}}$-component and for the $\hat{\boldsymbol{\theta}}$-component we will have:

\begin{equation}
r\frac{d^{2}\varphi}{dt^{2}}+{2}\frac{dr}{dt}\frac{d\varphi}{dt}=-\frac{1}{r}\frac{d\Phi}{d\theta}.\label{eqnmotion2}
\end{equation}

Now, taking equation (\ref{eqnmotion2}) and dividing throughout by $r\dot{\varphi}$ and remembering that the specific angular momentum $J=r^{2}\dot{\varphi}$, we will have:

\begin{equation}
\frac{1}{\dot{\varphi}}\frac{d\dot{\varphi}}{dt}+\frac{2}{r}\frac{dr}{dt}=-\frac{1}{J}\frac{d\Phi}{d\varphi}\Longrightarrow \frac{1}{J}\frac{dJ}{dt}= -\frac{1}{J}\frac{d\Phi}{d\theta},
\end{equation}

hence thus:

\begin{equation}
\frac{dJ}{dt}= -\frac{d\Phi}{d\theta}\label{jeqn}.
\end{equation}

The specific orbital angular momentum is the orbital angular momentum per unit mass and unless otherwise specified, we shall refer to it as angular momentum.

Digressing a little: what the above equation (\ref{jeqn}) means is that the orbital angular momentum of a planet around the Sun is not a conserved quantity. If it  is not conserved, then the sum of the orbital and spin angular momentum must be a conserved quantity (if this angular momentum is not say transfered to the Sun or other solar bodies), the meaning of which is that at the different $r$-positions, the spin of a planet about its own axis must vary. This could mean the length of the day must vary depending on the radial position away from the Sun. We shall come to this later, all we simple want to do is to underline this, as it points to the possibility of a secular change in the mean length of the day. 

Now moving on, if we make the transformation $u=1/r$, then for $\dot{r}$ and $\ddot{r}$ we will have:

\begin{equation}
\frac{dr}{dt}=-J\frac{du}{d\varphi}\,\,\textrm{and}\,\,\frac{d^{2}r}{dt^{2}}=-\frac{dJ}{dt}\frac{du}{d\varphi}-J^{2}u^{2}\frac{d^{2}u}{d\varphi^{2}},\label{tran}
\end{equation}

respectively. Inserting these into (\ref{eqnmotion1}) and then dividing the resultant equation by $-u^{2}J$ and remembering (\ref{jeqn}) and also that $dr=-du/u^{2}$, one is lead to:

\begin{equation}
\frac{d^{2}u}{d\varphi^{2}}-\left(\frac{1}{J^{2}u^{2}}\frac{d\Phi(u,\theta)}{d\varphi}\right)\frac{du}{d\varphi}+u=\frac{1}{J^{2}}\frac{d\Phi(u,\theta)}{du}.
\end{equation}

The solutions that we  shall consider  are those for which  $\theta$ is a time constant, \textit{i.e.} $r=r(\varphi)$  and for the convenience we shall write $\theta$ with subscript $p$, \textit{i.e.}, $\theta_{p}$. This is just to remind us that it ($\theta$) is not a variable in the equations of motion as this is a constant for a particular planet $p$, hence:

\begin{equation}
\frac{d^{2}u}{d\varphi^{2}}-\left(\frac{1}{J^{2}u^{2}}\frac{d\Phi(u,\theta_{p})}{d\theta_{p}}\right)\frac{du}{d\varphi}+u=\frac{1}{J^{2}}\frac{d\Phi(u,\theta_{p})}{du}\label{m1},
\end{equation}

and:

\begin{equation}
\frac{dJ}{dt}= -\frac{d\Phi(u,\theta_{p})}{d\theta_{p}}\label{m2}.
\end{equation}

This ends our derivation of the equations of motion for the field $\Phi(r,\theta)$. Before we proceed to our main task of showing how equations (\ref{m1} and \ref{m2}) explain the precession of planetary orbits, let us -- for instructive purposes, first lay down Einstein's solution.

\section{ Einstein's Solution}

When Einstein applied his newly discovered GTR to the problem of the precession of the perihelion of the planet mercury he obtained that the trajectory of solar planets must be described by the equation:

\begin{equation}
\frac{d^{2}u}{d\varphi^{2}}+u-\frac{G\mathcal{M}}{J^{2}}=\frac{{3}G\mathcal{M}u^{2}}{c^{2}}\label{euorb},
\end{equation}

where again $u=1/r$. To obtain a solution to this equation, we note that the left hand side is the usual Newtonian equation for the orbit of planets, \textit{i.e.}:
 
\begin{equation}
\frac{d^{2}u}{d\varphi^{2}}+u-\frac{G\mathcal{M}}{J^{2}}=0,
\end{equation}

and the solution to this equation is: $u=({1}+\epsilon\cos\varphi)/l$ where $\epsilon$ is the \textit{eccentricity} of the orbit and $l=({1}-\epsilon^{2})\mathcal{R}$ where $\mathcal{R}$ is half the size of the major axis of the ellipse. Written in different form, this solution is:

\begin{equation}
r=\left(\frac{1+\epsilon}{1+\epsilon\cos\theta}\right)\mathcal{R}_{min}.
\end{equation} 

where $\mathcal{R}_{min}$ is the planet's distance of closest approach to the Sun [see figure (\ref{pshift-fig}) for an illustration]. This solution is a good approximate solution to (\ref{euorb}) because the orbit of Mercury is nearly Newtonian. Consequently, we can rewrite the small term on the right hand side of (\ref{euorb}) as: ${3}G\mathcal{M}({1}+\epsilon\cos\varphi)^{2}/l^{2}c^{2}$; and in so doing, we make an entirely negligible error. With this substitution (\ref{euorb}) becomes:

\begin{equation}
\frac{d^{2}u}{d\varphi^{2}}+u-\frac{G\mathcal{M}}{J^{2}}=\frac{{3}G\mathcal{M}}{l^{2}c^{2}}\left({1}+{2}\epsilon\cos\varphi+\epsilon^{2}\cos^{2}\varphi\right)\label{euorb1}.
\end{equation}
 
and the solution to this equation is:

\begin{equation}
u=\frac{{1}+\epsilon\cos\varphi}{l}+\frac{{3}G\mathcal{M}}{l^{2}c^{2}}\left[{1}+\frac{\epsilon^{2}}{{2}}+\frac{\epsilon^{2}\cos{2}\varphi}{{{6}}}+\epsilon\varphi\sin\varphi\right],
\end{equation}

Of the additional terms, the first \textit{i.e.} ($1+\epsilon^{2}/2$) is a constant and the second oscillates through two cycles on each orbit; both these terms are immeasurably small. However, the last term increases steadily in amplitude with $\varphi$, and hence with time, whilst oscillating through one cycle per orbit; clearly this term is responsible for the precession of the perihelion. Dropping all unimportant terms we will have:

\begin{equation}
u=\frac{{1}+\epsilon\cos\varphi+\epsilon\eta\varphi\sin\varphi}{l},
\end{equation}

where $\eta={3}G\mathcal{M}/lc^{2}$ is extremely small. Thus all this leads us to:

\begin{equation}
u=\frac{1+\epsilon\cos\left(\beta_{E}\varphi\right)}{l},
\end{equation}

where: $\beta_{E}=\left({1}-\eta\right)$. At the perihelion we will have: $\beta_{E}\varphi={2}n\pi$ and this implies: $\varphi={2}n\pi\beta_{E}^{-1}\simeq{2}n\pi+{6}n\pi G\mathcal{M}/lc^{2}$. Essentially, this means that the perihelion advances by $\Delta\varphi={6}\pi G\mathcal{M}/lc^{2}$ per revolution and the resultant equation for the orbit is:

\begin{equation}
r=\frac{l}{1+\epsilon\cos\left(\varphi+\Delta\varphi\right)},
\end{equation}

hence thus the rate of precession of the perihelion is given by:

\begin{equation}
\left<\frac{\Delta\varphi}{\tau}\right>_{E}= \frac{{6}\pi G\mathcal{M}}{\tau c^{2}({1}-\epsilon^{2})\mathcal{R}}.
\end{equation}

This is Einstein's formula derived in $1916$ soon after he discovered the GTR. He [Einstein] concluded in the reading containing this formula:

\begin{quotation}
{``Calculation gives for the planet Mercury a rotation of the orbit of $43\arcsec$ per century, corresponding exactly to the astronomical observation (Leverrier); for the astronomers have discovered in the motion of the perihelion of this planet, after allowing for disturbances by the other planets, an inexplicable remainder of this magnitude. ''}
\end{quotation}

\begin{table*}
\begin{minipage}{126mm}
\begin{equation}
\Phi(u,\theta)=-G\mathcal{M}u\left[1+\lambda_{1}\left(\frac{G\mathcal{M}u}{c^{2}}\right)\cos\theta+\lambda_{2}\left(\frac{ G\mathcal{M}u}{c^{2}}\right)^{2}\left(\frac{{3}\cos^{2}\theta-{1}}{{2}}\right)\right].\label{orb0}
\end{equation}

\begin{equation}
\frac{d^{2}u}{d\varphi^{2}}+\left(\frac{\dot{J}}{J^{2}u^{2}}\right)\frac{du}{d\varphi}+u=-G\mathcal{M}u^{2}\left[{1}+\lambda_{1}\left(\frac{{2}G\mathcal{M}u\cos\theta}{c^{2}}\right)+\lambda_{2}\left(\frac{{3}G\mathcal{M}u}{c^{2}}\right)^{2}\left(\frac{{3}\cos^{2}\theta-{1}}{{2}}\right)\right],\label{orb1}
\end{equation}
\end{minipage}
\end{table*}

\section{\label{astgsol} Solution from the ASTG}

For the present, we shall take the second order approximation of the potential $\Phi(r,\theta)$ in-order to make our calculation for the precession of the perihelion of planetary orbits and this potential has been written down in (\ref{orb0}). As has already been said; we shall consider only those solutions for which  $\theta$ is a time constant, \textit{i.e.} $r=r(\varphi)$ and for the convenience that we do not think of $\theta$ as a variable we  have set $\theta:=\theta_{p}$. The solutions $r=r(\varphi)$ are those solutions for which the orbit of a planet stays put in the same $\theta$-plane. Now from the potential (\ref{orb0}) we shall have:

\begin{equation}
\frac{dJ}{dt}=-\left(\frac{G\mathcal{M}u}{c}\right)^{2}\left[\lambda_{1}\sin\theta_{p}+\lambda_{2}\left(\frac{{3}G\mathcal{M}\sin{2}\theta_{p}}{{2}rc^{2}}\right)\right].\label{varday}
\end{equation}
 
Now making the transformation $r={1}/u$, the first term on the left hand side of equation (\ref{orb1}) transforms to: 

\begin{equation}
\frac{d^{2}u}{d\varphi^{2}}+u-\frac{G\mathcal{M}}{J^{2}}=\beta_{1}u+\beta_{2}u^{2},\label{uorb}
\end{equation}

where:

\begin{equation}
\beta_{1}=\left(\frac{G\mathcal{M}}{J}\right)^{2}\left(\frac{{2}\lambda_{1}\cos\theta_{p}}{c^{2}}\right),
\end{equation}

and:

\begin{equation}
\beta_{2}l=\lambda_{2}\left(\frac{{3}G\mathcal{M}}{c^{4}}\right)\left(\frac{G\mathcal{M}}{J}\right)^{2}\left(\frac{{3}\cos^{2}\theta_{p}-{1}}{\textit{2}}\right)
\end{equation}

The left hand side of this equation (\textit{i.e.} \ref{uorb}) is what one gets from pure Newtonian theory and the term on the right is the new term due to the first order term in the corrected Newtonian potential and likewise the second term on the right is a new term due to the second order term in the corrected Newtonian potential. 

Now, taking the term $\beta_{1}u$ in equation (\ref{uorb}) to the right hand side, we will have:

\begin{equation}
\frac{d^{2}u}{d\varphi^{2}}+({1}-\beta_{1})u-\frac{G\mathcal{M}}{J^{2}}=\beta_{2}lu^{2}\label{orbapprox}.
\end{equation}

We know that the solution of the right hand side of the above equation when set to zero, \textit{i.e.}: 

\begin{equation}
\frac{d^{2}u}{d\varphi^{2}}+({1}-\beta_{1})u-\frac{G\mathcal{M}}{J^{2}}=0,
\end{equation}

is given by:

\begin{equation}
r=\frac{l}{{1}+\epsilon\cos(\eta_{1}\varphi)},\label{orbit}
\end{equation}

where:

\begin{equation}
\eta_{1}=\sqrt{{1}-\beta_{1}}=\sqrt{{1}-\left(\frac{G\mathcal{M}}{J}\right)^{2}\left(\frac{{2}\lambda_{1}\cos\theta_{p}}{c^{2}}\right)}.
\end{equation}

To obtain a  solution to (\ref{orbapprox}) to first order approximation, we note that the left hand side  has solution (\ref{orbit}) and that for nearly Newtonian orbits this solution $u=({1}+\epsilon\cos\varphi)/l$, is a good approximation to (\ref{orbapprox}) for nearly Newtonian orbits such as Mercury for example. Consequently, we can rewrite the small term on the right hand side of (\ref{orbapprox}) as: ${3}G\mathcal{M}({1}+\epsilon\cos\varphi)^{2}/l^{2}$; and make an entirely negligible error (see \textit{e.g.} \citealt{kenyon90}). With this substitution, equation (\ref{orbapprox}) becomes:

\begin{equation}
\frac{d^{2}u}{d\varphi^{2}}+\eta_{1}^{2}u-\frac{G\mathcal{M}}{J^{2}}=\frac{\beta_{2}}{l}\left({1}+{2}\epsilon\cos\varphi+\epsilon^{2}\cos^{2}\varphi\right),
\end{equation}
 
and the solution to this equation is:

\begin{equation}
u=\frac{{1}+\epsilon\cos\eta_{1}\varphi}{l}+\frac{\beta_{2}}{l}\left[\left({1}+\frac{\epsilon^{2}}{\textit{2}}\right)+\frac{\epsilon^{2}\cos{2}\varphi}{6}+\epsilon\varphi\sin\varphi\right].
\end{equation}

As before, \textit{i.e.}, as in the steps leading to Einstein's solution; of the additional terms, the first is a constant and the second oscillates through two cycles on each orbit; both these terms are immeasurably small. However, the last term increases steadily in amplitude with $\varphi$, and hence with time, whilst oscillating through one cycle per orbit; clearly this term is responsible for the precession of the perihelion. Now, dropping all the unimportant terms one is lead to:

\begin{equation}
u=\frac{{1}+\epsilon\cos\eta_{1}\varphi+\epsilon\eta_{2}\varphi\sin\eta_{1}\varphi}{l},
\end{equation}

where for the convenience we have set $\eta_{2}=\beta_{2}$ and this quantity is extremely small, in which case $\cos\eta_{2}\varphi\simeq1$ and $\sin\eta_{2}\varphi\simeq\eta_{2}\varphi$ and using these approximations (in the cosine addition formula $\cos\eta_{1}\varphi+\eta_{2}\varphi\sin\eta_{1}\varphi\simeq \cos\eta_{2}\varphi\cos\eta_{1}\varphi+\sin\eta_{2}\varphi\sin\eta_{1}\varphi=\cos\left[(\eta_{1}+\eta_{2})\varphi\right]$), we will have:

\begin{equation}
u=\frac{{1}+\epsilon\cos\left[(\eta_{1}+\eta_{2})\varphi\right]}{l}.
\end{equation}

Now, at the perihelion we are going to  have: $(\eta_{1}+\eta_{2})\varphi=2n\pi$ where $n={1},{2},{3}, \dots$ and this implies $\varphi={2}\pi n\left(\eta_{1}+\eta_{2}\right)^{-1}=2\pi n[\sqrt{1-\beta_{1}}+\beta_{2}]^{-1}\simeq2\pi n[1-(\beta_{1}-2\beta_{2})/2]^{-1}=2\pi n[1+(\beta_{1}/2-\beta_{2}) + ...]$ hence: $\varphi\simeq{2}\pi n+n\lambda_{1}h_{1}+n\lambda_{2}h_{2}$, where:

\begin{equation}
h_{1}=\left(\frac{{6}\pi  G\mathcal{M}}{lc^{2}}\right)\left(\frac{\cos\theta_{p}}{3}\right),
\end{equation}

and:

\begin{equation}
h_{2}=-\left(\frac{{3}\cos^{2}\theta_{p}-{1}}{{12}\pi }\right)\left(\frac{{6}\pi  G\mathcal{M}}{lc^{2}}\right)^{2}.
\end{equation}

This shows that per every revolution, the perihelion advances by:

\[
\frac{\Delta\varphi}{\tau}=\left(\frac{{6}\pi  G\mathcal{M}}{lc^{2}}\right)\left(\frac{\lambda_{1}\cos\theta_{p}}{3}\right)-\lambda_{2}\left(\frac{{3}\cos^{2}\theta_{p}-{1}}{{12}\pi }\right)\left(\frac{{6}\pi  G\mathcal{M}}{lc^{2}}\right)^{2},
\]

and this can be written more neatly and conveniently as:

\begin{equation}
\left<\frac{\Delta\varphi}{\tau}\right>_{O}= \left<\frac{\Delta\varphi}{\tau}\right>_{E}\left[\frac{\cos\theta_{p}}{{3}}\lambda_{1}-\left<\frac{\Delta\varphi}{\tau}\right>_{E}\frac{{3}\cos^{2}\theta_{p}-{1}}{{12}\pi\tau^{-1}}\lambda_{2}\right].\label{perishift}
\end{equation}

This formula -- which is a second order approximation; tells us of the perihelion shift of the planets. In the next section we will use this to deduce an estimate of the values of $\lambda_{1}$ and $\lambda_{2}$ for the Solar System and thereafter proceed to calculate the predicted values of the perihelion shift. As a way of showing that  these are solar values, let us denote  ($\lambda_{1}$ and $\lambda_{2}$) as  ($\lambda_{1}^{\tiny \odot}$ and $\lambda_{2}^{\tiny \odot}$) respectively.

\section{ An Estimate for $\lambda_{1}^{\tiny \odot}$ and $\lambda_{2}^{\tiny \odot}$}

If $\mathscr{P}_{p}$ is the precession per century of the perihelion of planet $p$, \textit{i.e.}: 

\begin{equation}
\mathscr{P}_{p}=\left<\frac{\Delta\varphi}{\tau}\right>_{O},
\end{equation}

then equation (\ref{perishift}) can be written as:

\begin{equation}
\mathscr{P}_{p}=\mathscr{A}_{p}\lambda^{\tiny \odot}_{1}+\mathscr{B}_{p}\lambda^{\tiny \odot}_{2},\label{simul}
\end{equation}

where:

\begin{equation}
\mathscr{A}_{p}= \left<\frac{\Delta\varphi}{\tau}\right>_{E}\left(\frac{\cos\theta_{p}}{{3}}\right),
\end{equation}

and:

\begin{equation}
\mathscr{B}_{p}= -\left(\frac{{3}\cos^{2}\theta_{p}-{1}}{{12}\pi\tau^{-1}}\right)\left(\left<\frac{\Delta\varphi}{\tau}\right>_{E}\right)^{2}.
\end{equation}

Given a set of the observed values for the size ($l_{p}$), the period of revolution $\tau_{p}$, the tilt ($\theta_{p}$) and the known precessional values of the perihelion of  planets ($\mathscr{P}^{obs}_{p}$); these values are listed in columns ${2}$, ${3}$, ${4}$ and ${8}$ of table (\ref{pshift}) respectively; we can solve for $\lambda^{\tiny \odot}_{1}$ and $\lambda^{\tiny \odot}_{2}$ since $\mathscr{P}_{p}$, $\mathscr{A}_{p}$ and $\mathscr{B}_{p}$ will all be known, thus one simple has to solve equation (\ref{simul}) for any pair of planets as a simultaneous equation.

The values of $\mathscr{A}_{p}$ and $\mathscr{B}_{p}$ for all the solar planets are listed in columns ${6}$ and ${7}$ of table (\ref{pshift}) respectively. It is important that we state that the values of the Inclination listed in column $4$ of table (\ref{pshift}) are the inclination of the planetary orbits relative to the ecliptic plane and in-order to compute the inclination of these orbits relative to the solar equator we have to add ${7}\degree$ to this because the ecliptic plane and the solar equator are subtended at this angle. The solar equator is here defined as the plane cutting the Sun into hemispheres and this plane is normal to the spin axis of the Sun.
\begin{table*}

\begin{minipage}{140mm}

\begin{center}
\textsc{\textbf{Perihelion Precession  of Solar Planetary Orbits According to the ASGT  }}
\\
\begin{tabular}{l r r r c l l r r c}
\hline\hline\\
\multicolumn{7}{c}{}&\multicolumn{3}{c}{\textbf{Precession}\,\,\,($1\arcsec/100{yrs}$)}\\
\multicolumn{7}{c}{}&\multicolumn{3}{c}{\textbf{{---------------------------------------------------------------------}}}\\
\textbf{Planet} & \multicolumn{1}{c}{$^{(b)}l_{p}$} & \multicolumn{1}{c}{$^{(b)}\tau_{p}$} & $^{(b)}$Incl.  & $^{(b)}$ $\epsilon$ & $\mathscr{A}_{p}$ & $\mathscr{B}_{p}$ & $\mathscr{P}_{p}^{obs}$ & $\mathscr{P}^{E}_{p}$ & $\mathscr{P}_{p}$ \\
 & $(\textrm{AU})$  & $({yrs})$ & ($\degree$) &  & & &  \\
\hline\\
{\textbf{Mercury}}  & ${0.39}$ & ${0.24}$ & ${7.0}$  & ${0.206}$  & ${3.50}\times{10}^{0}$ & ${1.72}\times{10}^{2}$ & $43.1000\pm0.5000^{(c)}$ & $43.50000$ & $42.80000\pm0.10000$ \\
{\textbf{Venus}}    & ${0.72}$ & ${0.62}$ & ${3.4}$  & ${0.007}$ & ${5.19}\times{10}^{-1}$ & ${2.88}\times{10}^{1}$ & $8.0000\pm5.0000^{(c)}$  & $\,\,\,8.62000$  & $12.00000\pm3.00000$\\
{\textbf{Earth}}    & ${1.00}$ & ${1.00}$ & ${0.0}$  & ${0.017}$ & ${1.57}\times{10}^{-1}$ & ${3.80}\times{10}^{-1}$ & $5.0000\pm1.0000^{(c)}$  & $\,\,\,3.87000$  & $\,\,\,4.00000\pm1.00000$\\
{\textbf{Mars}}     & ${1.52}$ & ${1.88}$ & ${1.9}$  & ${0.093}$  & ${7.02}\times{10}^{-2}$ & ${2.43}\times{10}^{-2}$ & $1.3624\pm0.0005^{(e)}$ & $\,\,\,1.36000$ & $\,\,\,1.70000\pm0.50000$\\
{\textbf{Jupiter}} & ${5.20}$ & ${11.86}$ & ${1.3}$ & ${0.048}$ & ${3.02}\times{10}^{-3}$ & ${1.00}\times{10}^{-5}$ & $0.0700\pm0.0040^{(e)}$ & $\,\,\,0.06280$ & $\,\,\,0.07000\pm0.02000$\\
{\textbf{Saturn}} & ${9.54}$ & ${29.46}$ & ${2.5}$ & ${0.056}$ & ${7.59}\times{10}^{-4}$ & ${1.72}\times{10}^{-7}$ & $0.0140\pm0.0020^{(e)}$ & $\,\,\,0.01380$ & $\,\,\,0.01900\pm0.00050$\\
{\textbf{Uranus}} & ${19.2}$ & ${84.10}$ & ${0.8}$ & ${0.046}$& ${1.09}\times{10}^{-4}$ & ${9.76}\times{10}^{-5}$ & $--^{(f)}$ & $\,\,\,0.00240$ & $\,\,\,0.00250\pm0.00070$\\
{\textbf{Neptune}} & ${30.1}$ & ${164.80}$ & ${1.8}$ & ${0.009}$ & ${3.98}\times{10}^{-5}$ &  ${9.13}\times{10}^{-11}$ & $--^{(f)}$ & $\,\,\,0.00078$ & $\,\,\,0.00270\pm0.00070$\\
{\textbf{Pluto}}$^{(a)}$ & ${39.4}$ & ${247.70}$ & ${17.2}$ & ${0.250}$ & ${5.77}\times{10}^{-5}$ & ${9.48}\times{10}^{-12}$ & $--^{(f)}$ & $\,\,\,0.00042$ & $\,\,\,0.00140\pm0.00040$\\

\hline\hline\\
\end{tabular}
\end{center}

\textbf{\underline{Notes}:}\,
$^{\textbf{(a)}}$ At the {2006} annual meeting of the International Astronomical Union, it was democratically decided that the solar test body Pluto is not a planet but a dwarf planet. For our purpose, its inclusion here as a planet is not affected by this decision for as long as this test body orbits the Sun like other planets.\,
$^{\textbf{(b)}}$ The values of $l_{p},\tau_{p},$ Inc. and Ecc. are adapted from \cite{sagan74}.\,
$^{\textbf{(c)}}$ Adapted from \cite{kenyon90}.\,
$^{\textbf{(d)}}$ Adapted from \cite{pitjeva05}.\,
$^{\textbf{(e)}}$ Obtained by adding the extra precession determined by \cite{pitjeva05} and found in \cite{iorio08b} to the standard Einsteinian perihelion precession.\,
$^{\textbf{(f)}}$ Because of their long orbital duration covering at least $2$ human lifetimes, no data is currently available covering one full orbital revolution for Neptune and Pluto hence there is not yet any observational values for the precession of their perihelia. The data for Uranus is unreliable (see \textit{e.g.} \citealt{iorio08b}).\, 

\caption[\textbf{Perihelion Precession of Solar Planetary Orbits According to the ASGT }]{\label{pshift} Above, column 1 gives the name of the planet $p$, column 2 gives $l_{p}$ which is the observed value for the orbital size of planet $p$, column 3 gives $\tau_{p}$ which is the period of revolution of the planet for one full orbit, column 4 is the tilt $\theta_{p}$  in degrees of the planet's orbit orbit relative to the ecliptic plane, column 5 gives the eccentricity of the orbit of the planet, while columns 6 and column 7 give the computed values $\mathscr{A}_{p}$ and $\mathscr{B}_{p}$ and  column 8,9 and 10 give   (1) the observed, (2) the GTR and (3) the ASTG precessional values of the planet.}
 
\end{minipage}
\end{table*}

Now, having calculated the values of  $\lambda^{\tiny \odot}_{1}$ and $\lambda^{\tiny \odot}_{2}$, we will have to use these values ($\lambda^{\tiny \odot}_{1}$ and $\lambda^{\tiny \odot}_{2}$) to check what are the predictions for the precession of the perihelion of the other seven planets. If the predictions of our theory are in agreement with the observed precession of the perihelion of these seven planets, then our theory is correct and if the predictions are otherwise then, our theory cannot be correct -- it must be wrong!

For the present, we have calculated $\lambda^{\tiny \odot}_{1}$ and $\lambda^{\tiny \odot}_{2}$ for the different planet pairs  were we have all the information to do so and these values are displayed in table (\ref{lambdacal}). The final adopted values are: 

\begin{equation}
\centering
\,\,\,\,\,\,\,\,\,\,\,\,\,\,\,\,\,\,\,\,\,\,\,\,\,\,\,\,\,\,\lambda_{1}^{\tiny \odot}={24.0\pm7.0} \,\, \textrm{and}\,\,\lambda_{2}^{\tiny \odot}={-0.200\pm0.100}.\label{lvals}
\end{equation}

and these values are the mean and the standard deviation -- there is a $27\%$ error in $\lambda_{1}^{\tiny \odot}$ and about twice ($50\%$) that error margin in $\lambda_{2}^{\tiny \odot}$. From the values given in (\ref{lvals}), the predicted values of the precession of the perihelion of the other seven planets \textit{i.e.} Earth, Mars, ..., Pluto; where computed and are listed in column {10} of table (\ref{pshift}). The equivalent predictions of these values from Einstein's theory are listed in column {9} of the same table. Inspection of the predictions of our theory reveals that our predicted values are -- as Einstein's predictions; in good agreement with observations. We believe that this does not mean the theory is correct but merely that it contains an element of truth in it. It means we have a reason to believe in it and as well a reason to peruse it further from the present exploration to its furthest reaches if this were at all possible!

The reader should take note that in our derivation, we have assumed as a first order approximation the Newtonian result namely that the angular momentum is a time constant. From the preceding section, clearly this is not the case. We have only assumed this as starting point of our exploration. It is hoped that taking into account the fact arising from the ASTG that orbital angular momentum is not a conserved quantity should lead to improved results that hopefully come closer to the observed values.

\begin{table}
\centering
\begin{minipage}{140mm}

\textsc{\textbf{ Estimation of the Values $\lambda_{1}^{\tiny \odot}$ and $\lambda_{2}^{\tiny \odot}$}}

\begin{tabular}{l l r}

\hline\hline
\textbf{Planet Plair}	& $\lambda_{1}$	& $\lambda_{2}$\\
	\hline
\textbf{Mercury-Venus}	    & $15.8$	& $-0.0716$ \\
\textbf{Mercury-Earth}	    & $32.8$	& $-0.4174$ \\
\textbf{Mercury-Mars}	      & $20.0$	& $-0.1574$ \\
\textbf{Mercury-Jupiter}	  & $26.1$	& $-0.2895$ \\ 
\textbf{Mercury-Saturn}	    & $27.6$  & $-0.3112$ \\
\hline
\textbf{Mean} &	\textbf{:}$\,24.0$	& $-0.200$\\
\textbf{Standard Deviation} &: 	$\,\,7.0$	& $-0.100$\\
\textbf{Percentage Error} &: 	$27\%$	& $50\%$\\
\hline\hline\\
\end{tabular}
\end{minipage}
\caption[\textbf{ Estimation of the Values $\lambda_{1}^{\tiny \odot}$ and $\lambda_{2}^{\tiny \odot}$}]{\label{lambdacal}  Column 1 in the first part of the table gives the name of the pair of the planets which have been used to obtain the pair of $\lambda$-values listed in columns 2 and 3. In the second part of the table we compute the Mean, Standard Deviation and Percentage Error of the $\lambda$-values}

\end{table}

\section{ None Conserved  Orbital Angular Momentum and its Implications}

Through equation (\ref{varday}) which clearly states that the orbital angular momentum of a planet must change with time; three immediate consequences of this are (1) a change in the mean Sun-Planet distance (2) a changing length of a planet's day  and (3) a secular change in solar spin. In the subsequent subsection, we shall go through these implied phenomena.

\subsection{Increase in Mean Sun-Planet Distance}

One of the most accurately determined physical parameters in astronomy is the mean Earth-Sun distance which is about the size of the Astronomical Unit ($AU$) where $1\,AU=149597870696.1\pm0.1m$ \citep{pitjeva05} and this is known to an accuracy of $10\,cm$ \citep{pitjeva05}. The Astronomical Unit according to the International Astronomical Union (Resolution No. 10 1976\footnote{see http://www.iau.org/static/resolutions/IAU1976 French.pdf}) is defined as the radius of an unperturbed circular orbit that a massless body would revolve about the Sun in $2\pi/k$ days where $k= 01720209895 AU^{3/2} day^{-1}$ is Gauss' constant. This definition is such that there is an equivalence
between the $\textrm{AU}$ and the mass of the Sun $\mathcal{M}_{\tiny \odot}$ which is given by $G\mathcal{M}_{\tiny \odot}=k^{2}A^{3}$. So, if $\mathcal{M}_{\tiny \odot}$ is fixed, it is technically incorrect to speak of a change $\textrm{AU}$. 

Before it was noticed that the mean Earth-Sun distance was changing it made perfect sense to refer to the mean Earth-Sun distance as the Astronomical Unit. Now, (1) because units must not change,  and (2) because of this fact that the mean Earth-Sun distance is changing; then, until\textsc{} such a time that the Astronomical Unit is correctly defined so that it is a true constant as physical unit must be,  it makes sense only to talk of the mean Earth-Sun distance instead of the Astronomical Unit. 

That the mean Earth-Sun distance is changing, this has been measured by \cite{krasinsky04} and \cite{standish05}. \cite{krasinsky04} finds $15.0\pm4.0\,m/cy$ which in SI units is  $(4.75\pm1.27)\times10^{-9}m/s$ and \cite{standish05} finds $7.00\pm0.20\,m/cy$ which in SI units is $(2.22\pm0.06)\times10^{-9}\,m/s$ where $1cy=100\,yr$.

To this rather surprising result, \textit{i.e.}, the apparent secular change in the mean Earth-Sun distance, \cite{iorio08a} states that the secular increase in the mean Earth-Sun distance can not be explained within the realm of classical physics. Contrary to this, we believe and hold that the ASTG can in-principle explain this result. The ASTG is well within the provinces of classical physics hence thus this result is explainable from within the domains and confines of classical physics. In his reading \citep{iorio08a} argues that the Dvali-Gabadadze-Porrati braneworld scenario -- a none-classical theory, which is a multi-dimensional model of gravity aimed to the explanation of the observed cosmic acceleration without darkenergy, predicts, among other things, a perihelion secular shift, due to Lue-Starkman Effect of $5 \times 10^{-4} arcsec/cy$ for all the planets of the Solar System. It yields a variation of about $6 m/cy$ for the increase in mean Earth-Sun distance; this is compatible with the observed time rate of change of the mean Earth-Sun distance hence giving the Dvali-Gabadadze-Porrati braneworld theory some breath. 

 \cite{iorio08a} goes on to say that the recently measured corrections to the secular motions of the perihelia of the inner planets of the Solar System are in agreement with the predicted value of the Lue-Starkman effect for Mercury, Mars and, at a slightly worse level, the Earth. We shall show that in-principle, the ASTG can explain this result as a consequence of the none-conservation of orbital angular momentum of planets in this azimuthally symmetric gravitational setting. The none-conversation of the orbital angular momentum leads directly to a time variation in the eccentricity of planetary orbits. This makes the secular change a purely classical result.

Now, given the definition of the eccentricity of an orbit: 

\begin{equation}
\epsilon^{2}=1-\left(\frac{\mathcal{R}_{min}}{\mathcal{R}_{max}}\right)^{2}
\end{equation}

where $\mathcal{R}_{min}$ and $\mathcal{R}_{max}$ are the spatial extent of the minor and major axis respectively; and then, differentiating this with respect to time, one is lead to: 

\begin{equation}
\epsilon\frac{d\epsilon}{dt}=-\frac{\mathcal{R}_{min}}{\mathcal{R}_{max}^{2}}\left(\frac{d\mathcal{R}_{min}}{dt}-\frac{\mathcal{R}_{min}}{\mathcal{R}_{max}}\frac{d\mathcal{R}_{max}}{dt}\right).
\end{equation}

There is no reason to assume that the rate of change of the minor and major axis be the same, thus we  must set:
 
\begin{equation}
\frac{d\mathcal{R}_{max}}{dt}=\left(\gamma+1\right)\left(\frac{d\mathcal{R}_{min}}{dt}\right),
\end{equation}

and from this it follows that:

\begin{equation}
\epsilon\frac{d\epsilon}{dt}=-\left(\frac{\mathcal{R}_{min}}{\mathcal{R}_{max}^{2}}\right)\left(1-\left(\gamma+1\right)\frac{\mathcal{R}_{min}}{\mathcal{R}_{max}}\right)\left(\frac{d\mathcal{R}_{min}}{dt}\right).
\end{equation}
 
and multiplying by $\mathcal{R}_{min}$ both sides, and thereafter substituting $\mathcal{R}_{min}/\mathcal{R}_{max}$ on the right hand side we will have:

\begin{equation}
\epsilon\mathcal{R}_{min}\frac{d\epsilon}{dt}=-(1-\epsilon^{2})\left(1-\left(\gamma+1\right)\sqrt{(1-\epsilon^{2})}\right)\frac{d\mathcal{R}_{min}}{dt},\end{equation}

therefore:

\begin{equation}
\frac{d\mathcal{R}_{min}}{dt}=-\frac{\mathcal{R}_{min}}{(1-\epsilon^{2})\left(1-\left(\gamma+1\right)\sqrt{1-\epsilon^{2}}\right)}\left(\epsilon\frac{d\epsilon}{dt}\right).
\end{equation}

Now, on the average, the time change of the minor axis must to a large extend be a good measure of the time change of the average distance $\left<\mathcal{R}\right>$ between the planet and the Sun, hence thus:

\begin{equation}
\frac{d\left<\mathcal{R}\right>}{dt}=\frac{\left<\mathcal{R}\right>}{(1-\epsilon^{2})\left(\left(\gamma+1\right)\sqrt{1-\epsilon^{2}}-1\right)}\left(\epsilon\frac{d\epsilon}{dt}\right).\label{evar}
\end{equation}

In the realm of Newtonian gravitation where spherical symmetry is assumed thus producing equations only dependent on the radial distance $r$, the eccentricity is an absolute time constant, \textit{i.e.}  $d\epsilon/dt\equiv0$, and this directly leads to  $d\left<\mathcal{R}\right>/dt\equiv0$, hence when one finds that the mean Earth-Sun distance is increasing, it comes more as a surprise. If we consider azimuthally symmetry in Poisson's equation as has been done here, the result emerges naturally because the eccentricity is expected to increase with the passage of time -- this we shall demonstrate very soon. 
 
In \S (4), against the clear message from the ASTG, we assumed that the orbital angular momentum of a planet is a conserved quantity. It turns out that taking this into account leads us to  two type of orbits (1) spiral orbits (2) the normal elliptical orbits with the important difference that the eccentricity of these orbits varies with time and it is this variation of eccentricity that we believe the secular increase of the mean Earth-Sun distance is rooted. 

Doing the right thing and taking into account the predicted change in the angular momentum, then equation (\ref{orbapprox}) will be:

\begin{equation}
\frac{d^{2}u}{d\varphi^{2}}+\left(\frac{1}{J^{2}u^{2}}\frac{dJ}{dt}\right)\frac{du}{d\varphi}+\eta_{1}^{2}u-\frac{G\mathcal{M}}{J^{2}}=\beta_{2}lu^{2},
\end{equation}

and taking the change of angular momentum to first order approximation from equation (\ref{varday}), one will have:

\begin{equation}
\frac{dJ}{dt}=-\left[\lambda_{1}\left(\frac{G\mathcal{M}}{c}\right)^{2}\sin\theta_{p}\right]u^{2}=-2\alpha u^{2},\label{varday2}
\end{equation}

where $\alpha$ is clearly defined from this equation \textit{i.e.}:

\begin{equation}
\alpha=\frac{1}{2}\left[\lambda_{1}\left(\frac{G\mathcal{M}}{c}\right)^{2}\sin\theta_{p}\right],\label{alpha}
\end{equation}

it therefore follows that:

\begin{equation}
\frac{d^{2}u}{d\varphi^{2}}-\frac{2\alpha}{J^{2}}\frac{du}{d\varphi}+\eta_{1}^{2}u-\frac{G\mathcal{M}}{J^{2}}=\beta_{2}lu^{2}.
\end{equation}

and writing $k=\alpha/J^{2}$ which is:

\begin{equation}
k=\frac{\lambda_{1}}{2}\left(\frac{G\mathcal{M}}{cJ}\right)^{2}\sin\theta_{p}=\frac{\lambda_{1}}{2}\left(\frac{G\mathcal{M}}{lc^{2}}\right)\sin\theta_{p}\label{k},
\end{equation}

where the Newtonian approximation $J^{2}=G\mathcal{M}l$ has been used and $K=G\mathcal{M}/J^{2}$, the above becomes:

\begin{equation}
\frac{d^{2}u}{d\varphi^{2}}-k\frac{du}{d\varphi}+\eta_{1}^{2}u-K=\beta_{2}lu^{2}.\label{neworb}
\end{equation}

If the orbital angular momentum varies constantly with time, then $J=\dot{J}t+J_{0}$ where $J_{0}$ is the angular momentum at time $t=0$ and $\dot{J}$ is a time constant, then $k=k(t)$ and $K=K(t)$, meaning $k(t)$ and $K(t)$ will dependent not on the coordinates $r,\theta,\varphi$ but only on time, hence in solving the above equation we can treat these as constants since they do not dependent on $r,\theta,\varphi$. We believe the assumption that $\dot{J}=constant$ is justified because if that was not the case, there could be an accelerated increase in the orbital angular momentum and this could have been noticed by now. In this assumption that  $\dot{J}=constant$, we must have $\dot{J}$  being so small that it is not easily noticeable as it appears to be the case since we have had to relay on delicate observations to deduce the secular increase of the mean Earth-Sun distance. 

Now, to obtain a solution to this equation (\textit{i.e.} \ref{neworb}), we need first to get a solution to:

\begin{equation}
\frac{d^{2}u}{d\varphi^{2}}-2k\frac{du}{d\varphi}+\eta_{1}^{2}u-\frac{G\mathcal{M}}{J^{2}}=0\label{neworb1},
\end{equation}

and to obtain a solution to this, we need first to solve:

\begin{equation}
\frac{d^{2}u}{d\varphi^{2}}-2k\frac{du}{d\varphi}+\eta_{1}^{2}u=0\label{orbtype2},
\end{equation}

and to its solution we add $G\mathcal{M}/J^{2}$. The axillary differential equation to this differential equation is: $X^{2}-2kX+\eta_{1}^{2}=0$ and its (\textit{i.e.} equation \ref{orbtype2}) solutions are:

\begin{equation}
X=k\pm \sqrt{k^{2}-\eta_{1}^{2}}=k\pm i\eta_{3},
\end{equation}

where $\eta_{3}=\sqrt{\eta_{1}^{2}-k^{2}}$. If $(\eta_{3})^{2}<0$ the solution is: $u=Ae^{\left(k+\eta_{3}\right)\varphi}+Be^{\left(k-\eta_{3}\right)\varphi}$ where $A$ and $B$ are constants, thus adding $G\mathcal{M}/J^{2}$ we have:

\begin{equation}
u=Ae^{\left(k+\eta_{3}\right)\varphi}+Be^{\left(k-\eta_{3}\right)\varphi}+\frac{G\mathcal{M}}{J^{2}}\label{sol1},
\end{equation}

and if $(\eta_{3})^{2}=0$ the solution is: $u=(A\varphi +B)e^{k\varphi}$ thus adding $G\mathcal{M}/J^{2}$ we have:

\begin{equation}
u=(A\varphi +B)e^{k\varphi}\label{sol2}+\frac{G\mathcal{M}}{J^{2}}.
\end{equation}

The solutions (\ref{sol1}) and (\ref{sol2}) are clearly spiral orbits. These solutions are obvious very interesting but because our focus is not on them, but on the solutions giving elliptical orbits in which the eccentricity varies, we shall not be looking into these spiral orbit solutions any further than we have already done.

Now, in the event that  $(\eta_{3})^{2}>0$  the solution to equation (\ref{neworb1}) is:

\begin{equation}
u=\frac{1+\epsilon e^{k\varphi}\cos(\eta_{3}\varphi)}{l}\label{sol3},
\end{equation}

Now, using the same strategy as that used in \S (3) and (4) to solving equations (\ref{euorb}) and (\ref{orbapprox}) respectively, one finds that the resultant orbit equation will be:

\begin{equation}
r=\frac{l}{1+\epsilon e^{k\varphi}\cos[(\eta_{2}+\eta_{3})\varphi]}\label{forb},
\end{equation}

and as before, at the perihelion we will have $(\eta_{2}+\eta_{3})\varphi=2\pi n$ and this implies $\varphi=2\pi n(\eta_{2}+\eta_{3})^{-1}\simeq2\pi n[\beta_{2}+ \sqrt{\eta_{1}^{2}-k^{2}}]^{-1}=2\pi n[\beta_{2}+ \sqrt{1-\beta_{1}-k^{2}}]^{-1}\simeq 2\pi n[1+(2\beta_{2}-\beta_{1})/2-k^{2}/2]^{-1}$ and taking only first order terms we will have: $\varphi\simeq2\pi n [1+(\beta_{1}-2\beta_{2})/2+k^{2}/2]$ and this shows that the perihelion will precess by an amount $\Delta \varphi=2\pi[(\beta_{1}/2-\beta_{2})+k^{2}/2]$, and in comparison with $\Delta\varphi\simeq2\pi [\beta_{1}/2-\beta_{2}]$ obtained without taking into account the changing angular momentum, there is an additional precession of $(\Delta\varphi)_{\textbf{+}}\simeq~\pi k^{2}$. The value of $k^{2}$ for the Solar System is so small that in practice, one can neglect it, thus, we have not missed out much in our calculation in which we have assumed a constant orbital angular momentum.  While this result is important our main thrust is to deduce the variation of the eccentricity of elliptical orbits (we shall shelf any deliberations on this result for a further reading). 

In equation (\ref{forb}), the term $\epsilon e^{k\varphi}$ in the denominator is the eccentricity, let us write this as $\epsilon_{*}=\epsilon e^{k\varphi}$, and from this we see that the eccentricity varies with time -- \textit{i.e.}; as the orbital angular momentum changes with the passage of time, so does the eccentricity. Now plucking this into equation (\ref{evar}) we can determine the variation of the mean Earth-Sun distance if we have knowledge of $\gamma$, \textit{unfortunately} we do not have this. However, if we are to reproduce the observed variation of the Earth-Sun distance, one finds that if they were to set $\gamma_{E}=1.48\times10^{-4}$, which practically means that the orbit grows evenly at every point, one is able to explain the secular increase of the mean Earth-Sun distance. 

It should be said that, if the ASTG is to stand on its own -- \textit{i.e.}, independent of observations, then it must be able to explain the result $\gamma_{E}=1.48\times10^{-4}$ from within its own provinces. It is for this reason that we say,  \textit{in-principle}, the ASTG is able to explain the secular increase in the mean Earth-Sun distance and only until such a time when one is able to derive say the value $\gamma_{E}=1.48\times10^{-4}$  from within the theory itself, will we be able to say the ASTG explains the secular increase in the Earth-Sun distance.

Other than the secular increase in the mean Earth-Sun distance, there is also the increase in the mean Earth-Moon distance. This has been measured by \cite{williams09} to be $\sim3.50\times10^{-3}\,m/cy$  and in SI units this is $1.11\times10^{-12}\,m/s$. This observation provides a test for the ASGT, but unfortunately, we do not have the value of  $\lambda_{1}$ so as to  check what the ASTG says about this. We believe one cannot use the same $\lambda$'s values obtained for the Sun because these values must be specific to the gravitating body and may very well be connected to the spin or the gravitating body in question. We are working on these ideas to improve the ASTG and at present we can only say it is prudent to assume that the $\lambda$-values are specific to the body in question hence one has to calculate them from observational data. For the Earth, this increase in the Earth-Moon distance is but the only observations we have in-order for us to deduce $\lambda_{1}^{\tiny{\earth}}$ hence the ASTG is unable to make any predictions on this as it stands in the present. We hope in the future one will be able to deduce a general form of the $\lambda_{\ell}$-values, thus placing the ASTG on a level where it is able to make predictions that are independent from observations. 

Important to note from $\epsilon_{*}=\epsilon e^{k\varphi}$ is that, as $\varphi\longmapsto-\infty$, the eccentricity will decrease and the reserve is that the eccentricity will increase as $\varphi\longmapsto+\infty$ decreases. An increasing eccentricity leads to a secular decrease in the Planet-Sun distance and a decreasing eccentricity leads to a secular increase in the Planet-Sun distance. This means the sense in which the planet orbits the Sun is important! Because we believe from \cite{krasinsky04} and \cite{standish05}, that there is a secular increase in the Earth-Sun distance, this means the current direction of rotation of the Earth around the Sun must be such that $\varphi\longmapsto-\infty$. This must be true for other planets rotating in the same sense as the Earth; and to any (object in the Solar System) that rotates in the direction opposite to this, this body will experience a secular decrease in its distance from the Sun.

\subsection{Secular Increase in the Orbital Period of Planets}

Given that through the passage of time -- what is suppose to be a sacrosanct parameter -- the mean Earth-Sun distance; is changing, and that the time change of the specific orbital angular momentum is given $\dot{J}=2r\dot{r}\dot{\theta}+r^{2}\ddot{\theta}$, then, if as in the case of Newtonian gravitation the specific orbital angular momentum of a planet is a conserved quantity,\textit{ i.e.} $\dot{J}=2r\dot{r}\dot{\theta}+r^{2}\ddot{\theta}=0$, then accompanying this result of a changing mean Earth-Sun distance must be an  increase in the  length of a planet's duration for one complete orbit since $\ddot{\theta}/\dot{\theta}=-2\dot{r}/r$. Given that $\dot{\theta}=2\pi/\mathcal{T}_{Y}$ where $\mathcal{T}_{Y}$ is the orbital period of a planet, the equation $\ddot{\theta}/\dot{\theta}=-2\dot{r}/r$ becomes: $\dot{\mathcal{T}}_{Y}/\mathcal{T}_{Y}=2\dot{r}/r$. Plucking in the relevant values for the Earth, one is lead to  $\dot{\mathcal{T}}^{\tiny{\earth}}_{Y}=2.97\, ms/cy$. Since $\mathcal{T}_{Y}^{\tiny{\earth}}=365.25\mathcal{T}_{D}$ where $\mathcal{T}_{D}^{\tiny{\earth}}$ is the period of an Earth day, it follows that: $\dot{\mathcal{T}}_{Y}^{\tiny{\earth}}=365.25\dot{\mathcal{T}}_{D}^{\tiny{\earth}}$,  further, follows that we must have: $\mathcal{T}_{D}^{\tiny{\earth}}=8.13\,\mu s/cy$ -- this value is at odds with physical reality; \textit{for} records held for over $2700\,yrs$ indicate that the Earth day changes by an amount $\dot{\mathcal{T}}^{\tiny{\earth}}_{D}=+1.70\pm0.05\, ms/cy$ (see \textit{e.g.} \citealt{miura09}), which is about $200$ times that expected if the orbital angular momentum where a conserved quantity as in Newtonian gravitation -- clearly, this suggests that the orbital angular momentum may not be conserved. 

If say the conserved quantity where the total angular momentum of a planet, \textit{i.e.} the sum total of the spin angular momentum ($S$) and the orbital angular momentum, then $\dot{S}=-\dot{J}$ and if the radius of the planet is not changing with time, then $\dot{\mathcal{T}}_{D}=-2\pi \mathcal{R}^{2}\dot{J}\mathcal{T}_{D}^{2}$. For the Earth, one finds that $\dot{\mathcal{T}}_{D}^{\tiny{\earth}}=-5.18\,s/cy$ which is $\sim3000$ times the observed value -- this can not be, sure something must be wrong. We shall explain this observational value $\dot{\mathcal{T}}^{\tiny{\earth}}_{D}=1.70\pm0.05\, ms/cy$ from the ASTG.

From the ASTG, we have:

\begin{equation}
\left(\frac{\dot{J}}{J}\right)_{\tiny{\earth}}^{theory}=-(6.00\pm2.00)\times10^{-15}s^{-1},
\end{equation}

and we know that:

\begin{equation}
\left(\frac{\dot{J}}{J}\right)_{p}=2\left(\frac{\dot{\mathcal{R}}}{\mathcal{R}}\right)_{p}-\left(\frac{\dot{\mathcal{T}_{Y}}}{\mathcal{T}_{Y}}\right)_{p}
\end{equation}

hence plucking in the observed values and remembering not to forget that for the Earth $\dot{\mathcal{T}}_{Y}^{\tiny{\earth}}=365.25\dot{\mathcal{T}}_{D}^{\tiny{\earth}}$, then we will have:

\begin{equation}
\left(\frac{\dot{J}}{J}\right)_{\tiny{\earth}}^{obs}=-(2.28\pm0.07)\times10^{-15}s^{-1}.
\end{equation}

This value -- \textit{vis}, the order of magnitude, is on a satisfactory level in good agreement with observations. We take this as further indication that the ASTG contains in it, a {grail} of the truth.

\subsection{Secular Increase in Solar Spin}

We know that angular momentum must be conserved but according to (\ref{varday}), it is not conserved. This lost orbital angular momentum must go somewhere -- it cannot just disappear into the thin interstices of spacetime or into the wilderness of spacetime thereof. Let $\mathcal{L}_{tot}$ be the sum total angular momentum of the Solar System, were we consider that the Solar System is composed of the planets. If the sum total of the angular momentum of a planet and its system of satellite is $J_{p}^{tot}$, then $\mathcal{L}_{tot}=\mathcal{M}_{\tiny \odot}S_{\tiny \odot}+\sum_{i}\mathcal{M}_{i}J_{i}^{tot}$. We would expect that the total angular momentum of the Solar System be conversed, that is $d\mathcal{L}_{tot}/dt=0$. From this we must have:

\begin{equation}
\frac{\dot{S}_{\tiny \odot}}{S_{\tiny \odot}}=-\frac{\dot{\mathcal{M}}_{\tiny \odot}}{\mathcal{M}_{\tiny \odot}}-\frac{1}{S_{\tiny \odot}}\sum_{i}\left[\frac{\mathcal{M}_{i}}{\mathcal{M}_{\tiny \odot}}\left(\frac{dJ_{i}^{tot}}{dt}\right)\right],
\end{equation}

and $dJ^{tot}_{p}/dt=dJ_{p}/dt$ hence thus:

\begin{equation}
\frac{\dot{\mathcal{T}}_{\tiny \odot}}{\mathcal{T}_{\tiny \odot}}=\frac{2\dot{\mathcal{R}}_{\tiny \odot}}{\mathcal{R}_{\tiny \odot}}+\frac{\dot{\mathcal{M}}_{\tiny \odot}}{\mathcal{M}_{\tiny \odot}}+\frac{\mathcal{T}_{\tiny \odot}}{2\pi \mathcal{R}_{\tiny \odot}^{2}}\sum_{i}\frac{\mathcal{M}_{i}}{\mathcal{M}_{\tiny \odot}}\frac{dJ_{i}}{dt},\label{sspin}
\end{equation}

and this means the orbital period of the Sun must be changing. If we assume that the Sun's radius has remained constant through the passage of time, \textit{i.e.} $\dot{\mathcal{R}}_{\tiny \odot}=0$ (which is certainly not true), then what we obtain from the above is a minimum value for the secular change in the Sun's spin. The reason for invoking this assumption is because there currently is no information on the secular change of the Sun's radius (see \textit{e.g.} \citealt{miura09}), hence we make this assumption so that we can proceed with our calculation. As already said, what we get is not the exact secular change in the Sun's spin but a lower limit to this.

The second term in equation (\ref{sspin}), \textit{i.e.} $\dot{\mathcal{M}}_{\tiny \odot}/\mathcal{M}_{\tiny \odot}$; represents the effect of solar mass loss, which can be evaluated in the following way. The Sun has a luminosity of at least $3.939\times10^{26}\,W$, or $4.382 \times 10^{9} \,kg/s$; this includes electromagnetic radiation and the contribution from neutrinos (\citealt{noerdlinger08}). The particle mass loss rate due to the solar wind is $\sim1.374\times10^{9}\, kg/s$ (see \textit{e.g.} \citealt{noerdlinger08}). From this information, it follows that $\dot{\mathcal{M}}_{\tiny \odot}/\mathcal{M}_{\tiny \odot}\simeq9.10\times10^{-12}cy^{-1}$.

Now, the last term in equation (\ref{sspin}) can be evaluated from the ASTG since $\dot{J}$ is known -- so doing, one finds that it is equal to $\sim-(4.00\pm1.00)\times10^{-6} cy^{-1}$; this implies $\dot{\mathcal{T}}_{\tiny \odot}=8.00\pm2.00\,s/cy$. This result is a significant $10^{6}$ times larger than the term emerging from the solar mass loss so much that we can neglect this altogether and consider only the last term in equation (\ref{sspin}) hence $\dot{\mathcal{T}}_{\tiny \odot}=8.00\pm2.00\,s/cy$. This value is significantly larger compared to that calculated by \cite{miura09} where these authors find a value of $21.0\,ms/cy$. Currently, no serious measurements on the secular change in the period of the solar spin has been made. It should be possible to undertake this effort and with respect to the ASTG, and the result of \cite{miura09}, this experiment would act an arbiter.

Furthermore, the authors \cite{miura09} propose that the Sun and the Earth are literally pushing each other away (leading to the increase in the Earth-Sun distance) due to their tidal interaction and they believe that this same process is what's gradually driving the moon's orbit outward: they say ``Tides raised by the moon in our oceans are gradually transferring Earth's rotational energy to lunar motion. As a consequence, each year the moon's orbit expands by about $4\, cm$ and Earth's rotation slows by about $30\mu s$''. Further \cite{miura09} assumes that our planet's mass is raising a tiny but sustained tidal bulge in the Sun. They calculate that, thanks to Earth, the Sun's rotation rate is slowing by $30 \mu s/cy$. Thus according to their explanation, the distance between the Earth and Sun is growing because the Sun is losing its angular momentum -- the ASTG gives a different explanation altogether and this is in our opinion, very interesting.

\section{ Discussion and Conclusions}

We have considered Poisson's equation for empty space  and solved this for an azimuthally symmetric setting -- we have coined the term Azimuthally Symmetric Theory \textit{of} Gravitation (ASTG) for the emergent theory thereof. From the emergent solution, we have shown that the ASTG is capable of explaining certain observed (and yet to be observed) anomalies:

\begin{enumerate}
\renewcommand{\theenumi}{(\arabic{enumi})}
\item Precession of the perihelion of planets.
\item Secular increase in the Earth-Sun distance.
\item Secular increase in the Earth Year.
\item Secular decrease in solar spin.
\item Spiral orbits must exist.
\end{enumerate}

One of the draw-backs of the ASTG as it currently stands is that it is heavily dependent on observations; \textit{for} the values of $\lambda_{\ell}$ need (have) to be determined from observations. Without knowledge of the $\lambda_{\ell}'s$, one is unable to produce the hard numbers required to make any quantifications. Clearly, a theory incapable of making any numerical quantifications is \textit{useless}. This must be averted. We shall make use of the solar values of the $\lambda_{\ell}'s$ in shading some light into our current thinking on this, \textit{i.e.} finding a general form for the constants $\lambda_{\ell}$; In the subsequent paragraphs, we shall make what we believe is a \textit{reasonable suggestion} and give our current envisage-ment on the general form for these constants.

(1) First things first, \textit{if} the constants $\lambda_{\ell}$ where all independent of each other, then, the theory would clearly be horribly complicated. If we take as guide the Principle \textit{of} Occam's Razor which in most if not all cases, leads to the simplest theory, then, these constants must be dependent on each other somehow so as to reduce the labyrinth of complications thereof. The simplest imaginable such dependence  is $\lambda_{\ell}=F(\ell)\lambda_{1}$; in this way, the entire system of constants $\lambda_{\ell}$ is dependent on just the one constant $\lambda_{1}$. This idea that the system of constants be dependent on just one constant is drawn from the theory of polynomial functions where for a  polynomial function $F(x)=\sum_{n=0}^{\infty} c_{n}x^{n}$, one can have ``well behaved'' polynomial functions for which the constants $c_{n}$ have a general form, \textit{i.e.}, were they dependent on $n$; \textit{e.g.}, $e^{x}=\sum_{n=0}^{\infty} x^{n}/n!$. We envisage the function $\Phi(r,\theta)$ to be a ``well behaved'' function. By ``well behaved'' we simple mean its system of constants, $\lambda_{\ell}$, is critically dependent on $\ell$ just as the constants $c_{n}$ depend on $n$.

(2) Second, we could like that on a practical level, only the second order approximation of the theory must suffice, this means the terms $\ell>3$ must be practically negligible. We have already shown herein  that the second order approximation of the ASTG is able to explain a sizable amount of anomalous observations. With the ASTG written in its second order approximation and as will be shown in the second reading (\textit{i.e.}, a follow-up reading that we hope will be published in the present journal), one is able without much difficulties, to explain from this second order approximation, the emergence of molecular bipolar outflows in star forming systems, as a gravitational phenomena. If the other terms beyond the second order approximation become practically significant, one will have difficulties to explain outflows. So in a way, we are not going to pretend but clearly state that, we want -- albeit with a priori and posteriori justification; to fine tune the theory so that it is able to explain the emergence of bipolar. This is the strongest reason we want the terms for which $\ell>3$ to be so small such that in practice one can neglect them entirely.

(3) Third and most important, the only data point we have of these constants is the determined values for the Sun, \textit{i.e.}, $\lambda_{\ell}^{\tiny \odot}=24.0\pm7$ and $\lambda_{2}=-0.2\pm0.1$. If logic is to hold -- as it must; then, our suggestion, $\lambda_{\ell}=F(\ell)\lambda_{1}$; must be able to explain this. We find that the following proposal:

\begin{equation}
\lambda_{\ell}=\left(\frac{(-1)^{\ell+1}}{\left(\ell^{\ell}\right)!\left(\ell^{\ell}\right)}\right)\lambda_{1},\label{lambda}
\end{equation}

meets (1), (2) and (3). We shall assume this result until such a time evidence to the contrary is brought forth. Checking on (3) we see that within the error margins $\lambda_{2}^{\tiny \odot}\simeq[(-1)^{2+1}/\left((2^{2})!(2^{2}\right)]\lambda_{1}^{\tiny \odot}$. Further checking on (2); from (\ref{lambda}) we will have $\lambda_{4}=3.40\times10^{-30}\lambda_{1}$ which is practically small; the meaning of which is that all terms for which $\ell>3$ can in practice be neglected entirely.

If the above proposal proves itself to be correct, then,  the resultant theory will have just one undetermined parameter $\lambda_{1}$. We are not going to try and deduce what this parameter depends on but simple hint at our current thinking. We believe this parameter must depend on the angular frequency of the spin of the gravitating body in question. If we can find the correct dependence, then, the ASTG will stand on it own thus positioning itself on the podium to make testable predictions. We have left the task to make this deduction an exercise for the follow-up reading.

The fact the we have deduced the crucial parameters $\lambda_{1}^{\tiny \odot}$ \& $\lambda_{2}^{\tiny \odot}$ from experience, means we have in the current reading done some reserve engineering. Normally, a theory must give these values and make clear predictions, just as when Einstein wrote down his equations and found that his theory predicted a factor $2$ difference when compared to Newton's theory when it come to the bending of light by the Sun and when applied to the Sun-Mercury system, it accounted very well for the then unexplained $43.0\arcsec$ per century for the precession of the perihelion of the orbit of this planet; it just came out right. There were no free parameters that needed fitting as is the case of the ASTG. As argued above, once a general form for the $\lambda_{\ell}$ is found, this setback of the ASTG will be solved. Because we were able to obtain the values $\lambda_{1}^{\tiny \odot}$ \& $\lambda_{2}^{\tiny \odot}$ which lead acceptable values for the perihelion precession, means that the values $\mathscr{A}$ \& $\mathscr{B}$ are not random but systematic. If the theory was all wrong, then, only luck would make the obtained values for $\mathscr{A}$ \& $\mathscr{B}$ give values of $\lambda_{1}^{\tiny \odot}$ \& $\lambda_{2}^{\tiny \odot}$ such that equation (\ref{simul}) give in general, acceptable values for the precession of the perihelion of the planets.

With regard to the values obtained for the precession of the perihelion of solar planets, it can be said that, the values obtained from the ASTG as shown in column  10 (table \ref{pshift}) when weighed against the observational values listed in column 10 (of the same table) are acceptable. Given that we have taken into account the fact the orbits of these planets are not found laying in the same plane, this can hardly be a coincident or an accident since changing their inclination by just $1\degree$ will alter the predicted values of the precession of their perihelion.

\cite{iorio08a} states that the secular increase in the mean Earth-Sun distance cannot be explained within the realm of classical physics. Contrary to this, we believe and hold that herein -- we have shown from within the provinces of classical physics that this result is explainable from within the domains and confines of classical physics. Before the present, the reason why perhaps this observation appeared beyond the reach of classical physics is because classical physics has not really considered gravitation as an azimuthally symmetric phenomenon as has been done in present reading. This strongly suggests that the ideas presented herein need to be explored further for they contain a debris of the truth.

One of the interesting outcome that was not explored in this reading for fear of digression is that the ASTG has a provision for spiral orbits (equation \ref{sol1} and \ref{sol2}). These orbits occur when $(n_{3})^{2}\leq0$. This condition implies the existence of a region ($r\leq\mathcal{R}_{crit}$) in which spiral orbits will occur. Evaluating the inequality $(n_{3})^{2}\leq0$, leads to: $\mathcal{R}_{crit}=(2\lambda_{1} G\mathcal{M}/c^{2})\cos^{2}(\theta/2)$, and from this, it is easy for one to deduce that spiral orbits are unlikely in the Solar System since these will have to occur inside the Sun because $\mathcal{R}_{crit}\leq\mathcal{R}_{\odot}$.

At this point as we approach the end of this reading, we feel strongly that we must address the question; ``Does the spin along the azimuthal axis of a gravitating body induce an azimuthal symmetry into the gravitational field for this spinning body?'' To answer this, we must ask the question; ``Will a contracting none-spinning cloud of gas experience any bulge alone its equator?'' First, we know that the equatorial bulge will occur on a plane perpendicular to the spin axis. Since a none-spinning gas cloud is going to have to spin axis, there is going to be no spin axis about which the equatorial bulge will occur. If the material in the cloud is randomly and uniformly distributed, the cloud will exhibit a spherically symmetric distribution of mass and its gravitational field is expected to be spherically symmetric. A spherically spherically symmetric gravitational field is one that only has a radial dependence, \textit{i.e.} $\varphi=\varphi(r)$. 

Now, if the gas cloud is spinning, the centrifugal forces will cause there to exist a disk and the material distribution will have an azimuthal symmetry, \textit{i.e}. $\rho=\rho(r,\theta)$. Should not this azimuthal symmetric distribution of matter induce an azimuthal gravitational field? From Poison's equation (\ref{poison}), $\rho=\rho(r,\theta)$ implies $\Phi=\Phi(r,\theta)$; should not this, \textit{i.e.} $\Phi=\Phi(r,\theta)$,  hold as-well for a body spinning gravitating body in a vacuum? From this, clearly, a spinning gravitating body ought to exhibit an azimuthal symmetry. We have argued in  the last paragraph of \S (2) that the spin of a gravitating body breaks the existing spherical symmetry of the non-spinning body and the above argument is just adding more to this. It is from this that the subtitle and running head finds its justification. 

If the ASTG turns out to be correct -- as we believe it will; then,  we have an important question  to ask; ``What is the speed of light doing in a theory of gravitation because from (\ref{bdef}) we see that the constants $G$ and $c$ are intimately tied-up together? This is a similar if not a congruent question that has been asked by \cite{martin09} in their expository work on Earth Flyby Anomalies (AFA). The empirical formula deduced to quantify EFA contains in it the speed of light, $c$, so in their exposition of the phenomena of AFA, \cite{martin09} have asked the perdurable question ``What is the speed of light doing there?''. EFA are thought to be a gravitational phenomena, so, what does the speed of light have to do with gravitation -- \textit{really}? If there is an intimate relationship between the speed of light and gravitation, then, one will be forgiven to think this suggests a link between gravitation and the theory of light -- electromagnetism. The speed of light, $c$, appears to be dire to the  ASTG presented herein.  Why not another value but the speed of light, $c$? We shall leave these matters hanging in-limbo.

In relation to the question above, \textit{i.e.}, ``What is the speed of light doing in a theory of gravitation'',  one notes that Newtonian gravitation -- which requires instantaneous interaction as a postulate; does not imply the dependence of the gravitational potential on the azimuthal angle for a spinning body as is the case in the ASTG, because at any instant $t$, the gravitating body appears  spherically symmetric. Here we  have the speed of light $c$ coming in because of the azimuthal symmetry. Does this speed of light $c$ link (or not) the propagation of the gravitational phenomena to the speed of light? In the present, we can only pause this as a question, for we still have to do further work on these ideas. 

In closing, allow me to say that we find it hard to call what has been presented herein as ``A New Theory \textit{of} Gravitation''. When one tells you they have come up with a new theory of gravitation, what immediately comes to mind is that they have discovered a new principle upon which gravitation can further be understood from the present understanding. The ASTG is not founded on any new physical principle but on the well known vintage  equation of Poisson. What we have done is simply taken the azimuthally symmetric equations of this equation and applied them to gravitation. Based on this understanding, it is difficult to call it a new theory. Yes, the azimuthally symmetric equations of Poisson have brought new and exciting physics -- perhaps only because of this, the title of this reading finds its qualification.

\textbf{\underline{Acknowledgments}:} I am grateful to  Mkoma George -- Baba \textit{va} Panashe, and his wife -- Mai \textit{va} Panashe, for their kind hospitality they offered while working on this reading and to Mr. Isak D. Davids \& Ms. M. Christina Eddington for proof reading the grammar and spelling. Further, I am grateful to the anonymous reviewers -- for their invaluable criticism that has helped in the refinement of the arguments presented. Last and certainly not least, I am very grateful to my Professor, D. Johan van der Walt and Professor Pienaar Kobus, for the strength and courage that they have given me.

\label{lastpage}
\end{document}